\documentclass{aa}
\begin{document}

\title{Distances of the TeV SNR complex CTB 37 towards the Galactic Bar} 

\author{W.W. Tian \inst{1}
\and
D.A. Leahy \inst{2}}
\authorrunning{Tian\& Leahy}
\titlerunning{Distances of the TeV SNR complex CTB 37}
\offprints{tww@bao.ac.cn}
\institute{National Astronomical Observatories, CAS, Beijing 100012, China;\\
\and
Department of Physics \& Astronomy, University of Calgary, Calgary, Alberta T2N 1N4, Canada}

\date{Accepted XX; Received 2012; in original form 2012 XX}

\abstract{
Three supernova remnants form the CTB 37 complex:
CTB 37A (G348.5+0.1, associated with the TeV $\gamma$-ray source HESS J1714-385), 
CTB 37B (G348.7+0.3, associated with HESS J1713-381 and the magnetar CXOU J171405.7.381031), 
and G348.5-0.0.  
We use 21 cm HI absorption measurements to constrain the kinematic distances to these SNRs, 
which have not previously been determined well. 
We revise the kinematic distance for CTB 37A to be in the range 6.3 to 9.5 kpc 
(previously $\sim$11.3 kpc) 
because it is beyond the near 3-kpc arm and in front of the far side of the CO cloud at -145 km~s$^{-1}$ towards $l$=348.5. 
G348.5-0.0 has a HI column density (N$_{HI}$ $\sim$6.1$\times$10$^{21}$ cm$^{-2}$) lower than 
CTB 37A ($\sim$7.1$\times$10$^{21}$ cm$^{-2}$). 
Also, G348.5-0.0 does not have the major absorption feature at -107 km~s$^{-1}$
that CTB 37A shows. 
This is caused by the near 3-kpc arm, so G348.5-0.0 is at a distance of $\le$ 6.3 kpc. 
CTB 37B is at a distance of $\sim$13.2 kpc (previously 5 to 9 kpc) based on: 
1) it has an absorption feature at -10$\pm$5 km~s$^{-1}$ from the far 3-kpc arm, so CTB 37B is behind it; 
2) there is absorption at -30 km~s$^{-1}$ but not at -26 km~s$^{-1}$, which yields the distance value; 
3) the HI column density towards CTB 37B ($\sim$8.3$\times$10$^{21}$ cm$^{-2}$) is larger than CTB 37A. 
In summary, CTB 37A, CTB 37B and G348.5+0.0 are all at different distances and are only by
chance nearby each other on the sky.
In addition, we conclude that CTB 37 A and B are not associated with the historical Supernova AD 393.   

}
\maketitle

\section{Introduction and Data}

\begin{figure}
\vspace{45mm}
\begin{picture}(50,50)
\put(-10,-90){\includegraphics{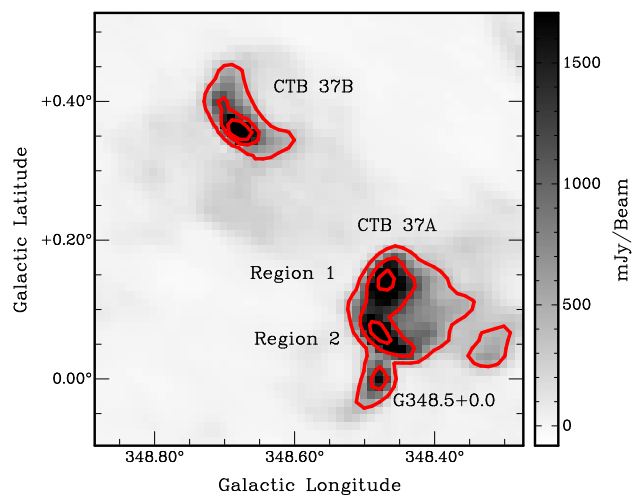}}
\end{picture}
\caption{SGPS 1420 MHz continuum image of SNRs CTB 37A/B and G348.5-0.0, with contours at 0.4, 1.2 and 2.1 Jy/Beam.}
\end{figure}

Distance measurements of Galactic supernova remnants (SNRs) are key to the study of SNRs. 
Distance is not only needed to find their properties such as radius, explosion energy, size, age and 
local interstellar medium density, 
but also help us understand related high-energy astrophysical phenomena. 
Recently, the TeV emission from SNRs has become of great interest, with new TeV $\gamma$-ray 
observations identifying some SNRs as TeV sources. 
Distance determination to TeV SNRs can help explore the mechanisms for TeV emission from SNRs 
(e.g. leptonic or hadronic origins). 

Most Galactic TeV sources are located in the Galactic disk where dust and gas 
leads to serious extinction in the optical, ultraviolet and X-ray bands. 
However the dust and gas are transparent to radio and infrared emission. 
It has been more than half century since 21cm HI absorption observations became a major tool 
to measure kinematic distances of radio objects in the disk. 
These are usually based on a circular rotation curve model of our Galaxy (e.g. Hou, Han \& Shi 2009; Brand \& Blitz 1993). 
We have carried out follow-up studies of TeV SNRs with strong radio emission 
by employing the 1420 MHz continuum data and the HI line data from the recently-finished International Galactic Plane Survey. 
Improved distance-measurement methods were used to find parameters of several TeV SNRs (i.e. W41/HESS J1834-087, Kes 75/HESS J1846-029, G21.5-0.9/HESS J1833-105, G54.1+0.3, SNR G353.6-0.7/HESS J1731-347, see Tian \& Leahy 2010; Tycho SN 1572, see Tian \& Leahy 2011). 
Our methods include two main improvements over previous methods (see Leahy \& Tian 2010 for details). 
In general, our methods have helped solve the near-far kinematic distance ambiguity in the inner Galaxy and the kinematic distance confusion due to the velocity reversal within the Perseus arm. 
Specially, many SNRs are interacting with CO clouds (Table 2 in Jiang et al. 2010), our methods are powerful to determine their distances (Tian, Leahy \& Li 2010).  

For the inner Galaxy (inside the 3 kpc molecular ring), it is generally known that a circular
 rotation curve model is not suitable. Inside 3 kpc, the gas likely follows oval orbits because
of the effect of Galactic bar potential. 
Several recent studies have shown that it is challenging to estimate distances along 
lines-of-sight towards the 3 kpc ring (i.e. Galactic longitude $|l|\sim<20^\circ$).
Thus previous distance estimates to some objects in these directions may have large uncertainty or even be incorrect. 
Recently-finished radio (CO and HI) and infra-red surveys reveal detailed kinematic images of gas and dust in the Galactic (GC) region, including the far 3 kpc arm's discovery (Dame \& Thaddeus 2008). 
Several gas flow models in the GC region have been presented based on recent observations (e.g. Rodriguez-Fernandez \& Combes 2008). 
This justifies our undertaking of a new study of distances with the new data for objects in the
inner Galaxy. In this paper, we provide revised distances to SNR G348.0-0.0 and TeV SNRs CTB 37A/B. 

We use the 1420 MHz radio continuum and 21 cm HI data from the Southern Galactic Plane Survey (SGPS), which was taken with the Australia Telescope Compact Array and the Parkes 64m single dish telescope (McClure-Griffiths et al. 2005, Haverkorn et al. 2006). 
The continuum observations have a resolution of 100~arcsec and a sensitivity better than 1~mJy/beam. 
The HI data have an angular resolution of 2~arcmin, a rms sensitivity of $\sim$1~K and a velocity resolution of 1~km~s$^{-1}$. 
The $^{12}$CO ($J$=1-0) data for CTB 37A and G348.0-0.0 is from survey data taken with the 12 m NRAO telescope (see Reynoso \& Mangum 2000 for details).

\section{Results and Analysis}

We show the 1420 MHz continuum image of the CTB 37 SNR complex in Figure 1. 
HI emission spectra for source and background regions, and the resulting
absorption spectrum were constructed.
Fig. 2 shows the results for four regions: two regions with bright continuum in CTB 37A 
(the two bright regions shown in Fig. 1); and one region each for CTB 37B and G348.5-0.0. 
The respective CO emission spectra for the same source regions in CTB 37A and G348.0-0.0
(CTB 37B is outside  of the CO survey data) 
are shown in the lower part of each panel, scaled vertically for easy comparison with
the HI absorption spectra. 
We have carefully analyzed the reality of absorption features in the spectra, including comparison with their respective CO emission spectra when CO data is available.

\begin{figure*}
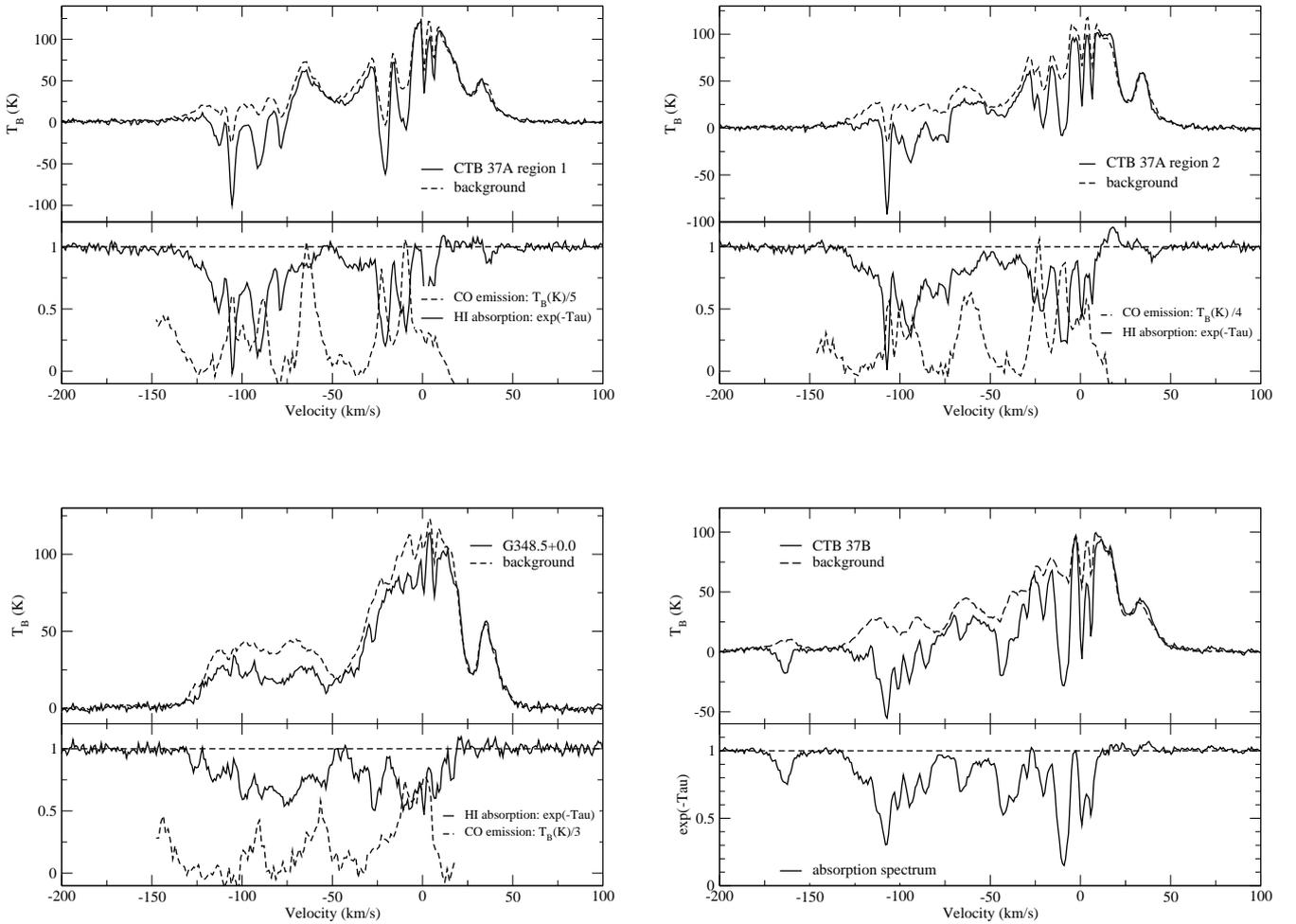
 
\vspace{110mm} 
\begin{picture}(80,80)
\put(0,180){\includegraphics{f2a.eps}}
\put(260,180){\includegraphics{f2b.eps}}
\put(0,-20){\includegraphics{f2c.eps}}
\put(260,-20){\includegraphics{f2d.eps}}
\end{picture}
\caption{HI emission spectra for source (solid line) and background (dashed line) regions are shown in the top half of each panel; HI absorption spectra and $^{12}$CO emission spectra for each source region are shown in the lower half of each panel.  The upper-left panel is for CTB 37A region 1; the upper-right panel is for CTB 37A region 2; 
the lower-left panel is for G348.5-0.0; and the lower-right panel is for CTB 37B.}
\end{figure*} 

CTB 37A's HI spectra are extracted from two bright regions (Regions 1 and 2 in Fig. 1). 
In Region 2, G348.5-0.0 may overlap CTB 37A but the bright emission should be from CTB 37A 
because it matches the shell structure of CTB 37A, and G348.5-0.0 has lower continuum brightness. 
The HI absorption spectrum from Region 2 has very similar structure to that from Region 1,
but is significantly different from that from G348.5+0.0, confirming that the continuum 
from G348.5+0.0 is small in region 2.  
Comparison of the HI absorption spectra with the respective CO spectra for CTB 37A and 
G348.5-0.0 shows that the CO peaks have associated HI absorption features except for the lowest velocity CO peak at -145 km s$^{-1}$. 
This shows that the CO cloud at -145 km s$^{-1}$ is behind both CTB 37A and G348.5-0.0. 
The CO peak at $\sim$ -65 km/s towards CTB 37A has an associated HI absorption feature with an optical depth of $\tau \sim$ 0.2, 
much smaller than other HI features ($\tau \ge$ 0.5) with associated CO peaks. 
Thus most CO at -65 km s$^{-1}$ is behind CTB 37A. 
The lowest velocities of HI absorption features towards CTB 37A and G348.5-0.0 are the same and are about -125 km s$^{-1}$. 
For both CTB 37A and G348.5-0.0, absorption features appear from local (10 to 15 km s$^{-1}$) up to the most negative velocity, 
but the HI column density of N$_{HI}\sim$6.1$\times$10$^{21}$ cm$^{-2}$ towards G348.5-0.0 is smaller than towards CTB 37A of N$_{HI}\sim$7.1$\times$10$^{21}$ cm$^{-2}$
(using N$_{HI}$=1.9$\times$10$^{18}\tau\Delta vT_{s}$, assuming $T_{s}$=100 K). 

The HI absorption spectrum of CTB 37B shows all major absorption features in CTB 37A's HI spectrum, 
but also has three additional features at -165 km s$^{-1}$, at -65 km s$^{-1}$, and at -30 to -45 km s$^{-1}$. 
The channel map at -165 km s$^{-1}$ (shown in Fig. 3) reveals a diffuse high velocity HI cold cloud (T$_{B}\le$ 20 K) covering CTB 37B. 
This cloud is responsible for the HI absorption feature in the CTB 37B HI spectrum, 
as further illustrated in the lower panel of Fig. 3 by the negative T$_{B}$ at the peak of the continuum emission. 
The HI column density of N$_{HI}$ $\sim$8.3$\times$10$^{21}$ cm$^{-2}$ towards CTB 37B is larger than towards CTB 37A. 
The above differences show that CTB 37B is more distant than CTB 37A and G348.5-0.0.   

\begin{figure}
\vspace{80mm}
\begin{picture}(80,80)
\put(-55,375){\includegraphics{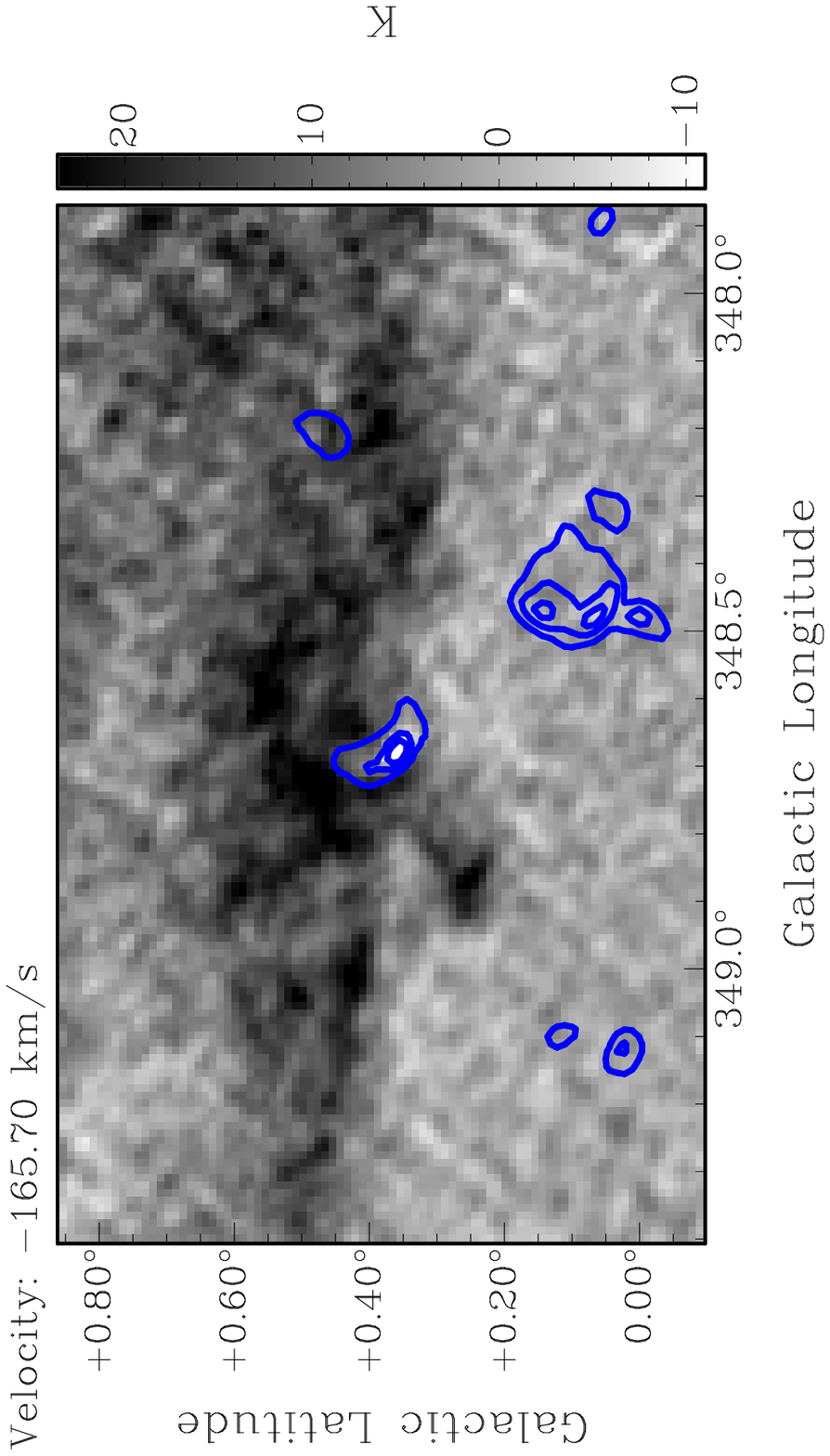}}
\put(-10,-80){\includegraphics{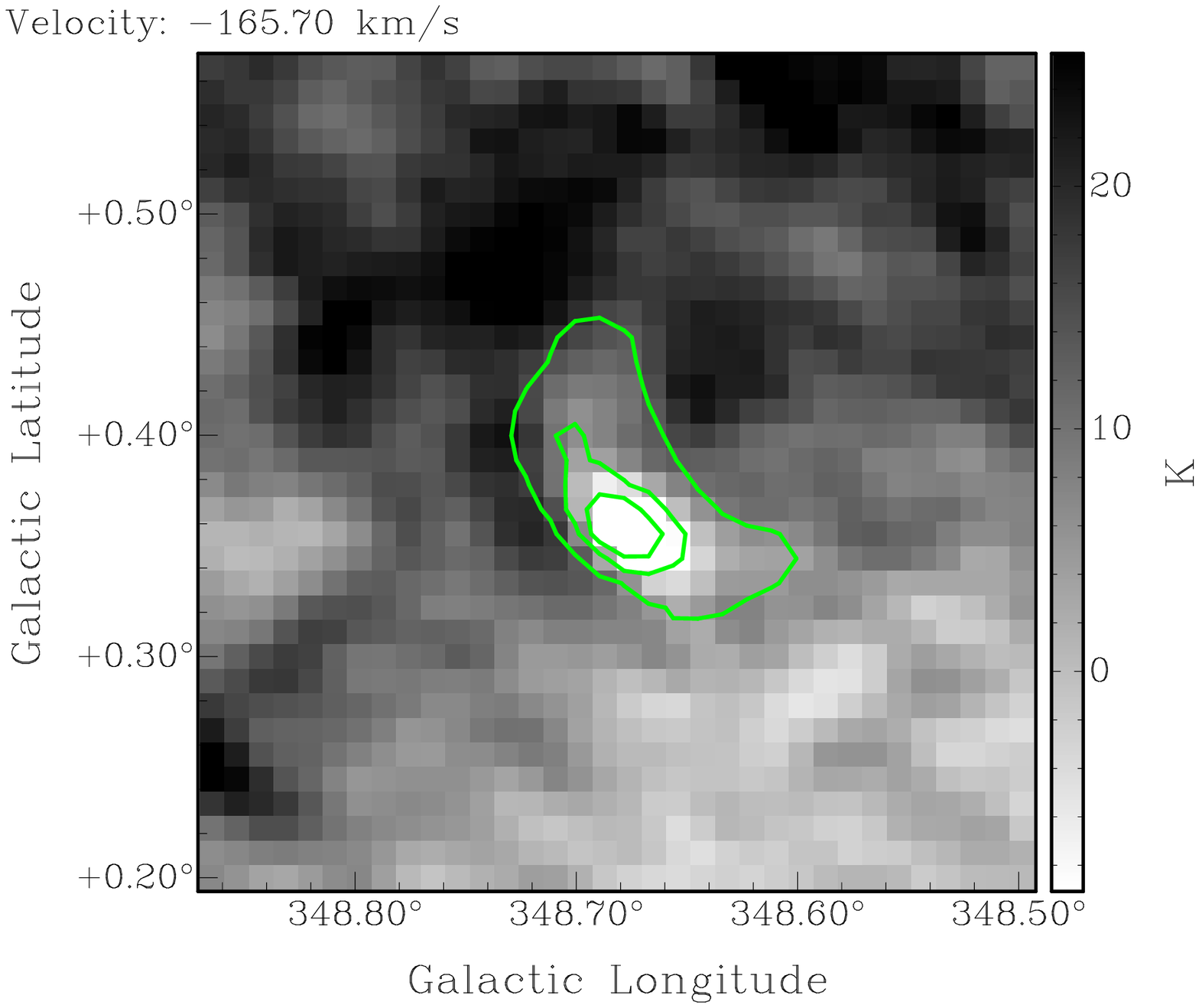}}
\end{picture}
\caption{SGPS HI channel map at -165.7 km/s (upper-large scale, lower-detail), with overlay
of 21cm continuum contours from Fig. 1.}
\end{figure}

\section{Discussion and Conclusion}  
\subsection{Distance}
21 cm HI absorption spectra can provide distance estimates to Galactic objects using 
a radial velocity-distance relation based on the circular rotation curve model. 
However, the circular rotation model fails when objects are in the inner Galaxy, 
because the gas follows non-circular orbits in the Galactic bar potential
(see Weiner \& Sellwood 1999). 
Recently Dame \& Thaddeus (2008)detected symmetric expanding 3 kpc arms in CO emission,
supporting the existence of a bar at the center of our Galaxy. 
The longitude-velocity diagram of CO emission for -12$^{0}<l<13^{0}$ given by Bitran et al. 
(1997) and Dame \& Thaddeus (2008) shows a central velocity towards $l$=348.5$^{0}$
of $\sim$-5 km~s$^{-1}$ for the far arm and $\sim$-101 km~s$^{-1}$ for the near arm,
and that the two arms have CO line widths of $\sim$ 21 km~s$^{-1}$. 
The velocity for the far arm is very different from the previous value of $\sim$ -50 km~s$^{-1}$ 
from the dispersion ring model (Shane 1972). 
The latter has been widely used to interpret HI spectra of objects towards the Galactic center and estimate kinematic distances (Caswell et al. 1975). 
Rodriguez-Fernandez \& Combes (2008) have simulated the gas dynamics in the inner Galaxy and gave a gas distribution model for the GC region. 
In the following section we use these models to explain the above HI and CO spectra and obtain revised distance estimates 
for CTB 37A/B and G348.5-0.0.

\subsubsection{Distance of CTB 37A/G348.5-0.0}
The HI spectrum of CTB 37A shows a major absorption peak at velocity of $\sim$ -107 km~s$^{-1}$, 
 consistent with the expected radial velocity of the near 3 kpc arm at $\l$=348$^{\circ}$ to $\sim 349^{\circ}$. 
This constrains the remnant's distance to d$\ge$ 6.3 kpc (taking the rotation curve model with R$_{o}$=8.5 kpc, V$_{o}$=220 km~s$^{-1}$ for R$\ge$ 3 kpc Galactic regions, and non-circular rotation for the 3 kpc arms). 
The terminal velocity at $l$=348$^{0}$ to 349$^{0}$ is -190 km~s$^{-1}$ (Fig. 2 of both Weiner \& Sellwood 1999 and Burton \& Liszt 1993), 
Yet the lowest emission in the spectrum of CTB 37A extracted from the SGPS data (with a 
sensitivity of 1.6 K) only reaches 125 km~s$^{-1}$, 
at which there is an associated weak HI absorption feature. 
However the CO emission towards CTB 37A shows a most negative velocity CO peak of -145 km~s$^{-1}$
 at which we do not detect any associated HI emission or absorption feature. 
This shows that the CO cloud is behind CTB 37A and G348.5+0.0.
The interstellar gas in the inner 3 kpc has asymmetric structure, and the -145 km~s$^{-1}$ CO
 cloud may be either at the near or far value of distance ($\sim$ 7.2 of 9.5 kpc) for -145 km~s$^{-1}$ (see Fig. 8 of Weiner \& Sellwood 1999). 
Therefore we obtain an upper distance limit of 9.5 kpc.

The optical depth of the absorption feature at -10$\pm$5 km~s$^{-1}$ is $\tau \sim$ 0.4 in 
CTB 37A compared to $\tau \sim$ 0.7 for CTB 37B. 
Yet the feature at -107 km~s$^{-1}$ almost has the same optical depth ($\sim$ 0.7) for both CTB 37A and 37B. 
Thus both CTB 37A/B are behind the near 3 kpc arm at -107 km~s$^{-1}$. 
The extra absorption at -10 km~s$^{-1}$ for CTB 37B implies that CTB 37B is behind the far 3-kpc arm at -10 km~s$^{-1}$. 
It is known (Dame \& Thaddeus 2008) that the near arm has more HI than the far arm consistent with the observed optical depth ($\tau \sim$ 0.7 for the near arm and $\tau \sim$ 0.3 for the far arm). 
Based on this, the absorption at -10 km~s$^{-1}$ in CTB 37A is likely caused by local gas.

The HI absorption spectrum of G348.5-0.0 does not show the major -107 km~s$^{-1}$ absorption feature. 
Thus the remnant is in front of the near 3-kpc arm, with an upper limit distance of 6.3 kpc. 
The CO spectrum of G348.5-0.0 also shows an peak emission at -145 km~s$^{-1}$ without associated HI emission and absorption features. 
This is consistent with the remnant being in front of the CO cloud. 
From the HI absorption spectra we calculate the HI column density of G348.5-0.0 (6.1 $\times$10$^{21}$ cm${-2}$) and of CTB 37A( 7.1$\times$10$^{21}$ cm${-2}$). This is
consistent with G348.5-0.0 being nearer than CTB 37A. 
 
\subsubsection{Distance of CTB 37B}
The HI absorption spectrum of CTB 37B shows all the major features of CTB 37A's absorption spectrum. 
CTB 37B shows additional features: most negative velocity absorption feature at -165 km~s$^{-1}$
 and obvious absorption peaks at -65, -45 and -30 km~s$^{-1}$. 
CTB 37B is more distant than CTB 37A, because it is behind the far 3-kpc arm. 
We note an emission peak at -26 km~s$^{-1}$ that has no respective absorption feature.
This gives therefore a best distance estimation of 13 to 13.4 kpc (not including systematic error here), 
i.e. CTB 37B behind far side of position with -30 km~s$^{-1}$ and in front of the far side 
position with -26 km~s$^{-1}$. 
In this case, the absorption features -65 and -45 km~s$^{-1}$ should be associated with  
cold HI clouds at the far side velocities (so they do not appear in CTB 37A's absorption spectrum). 
Figure 3 shows that the -165 km~s$^{-1}$ absorption feature is caused by a diffuse cold HI cloud 
which does not cover CTB 37A, about 0.3$^{\circ}$ away from CTB 37B. 
This cloud might be close to the dustlane inside the 3 kpc ring because of its large radial velocity (Fig. 10 of Rodriguez-Fernandez \& Combes 2008).   

\subsection{Comparing Our Measurements with the Previous Estimation}
\begin{table}
\begin{center}
\caption{Comparison of distance estimates from the past and present for SNR complex CTB 37}
\setlength{\tabcolsep}{1mm}
\begin{tabular}{cccc}
\hline
SNR Name: &  37A &  37B  & G348.5-0.0 \\
\hline
\hline the past:  & $\sim$11.3 kpc$^{1}$ &5 to 9 kpc$^{2}$ & $\sim$14 kpc$^{3}$ \\
Ours:  &  6.3 to 9.5 kpc & $\sim$13.2 kpc& $\le$ 6.3 kpc \\ \hline \hline
\end{tabular} \end{center}
References: $^{1}$Reynoso \& Mangum  2000; $^{2}$Aharonian et al. 2008, Caswell et al. 1975; $^{3}$Reynoso \& Mangum  2000.
\end{table}

Caswell et al. (1975) suggested that CTB 37A/B lie between 6.7 and 13.7 kpc based on the rotation curve model 
of Schmidt \& Blaauw (1965), because they detected the -107 km~s$^{-1}$ absorption feature and 
failed to detect any absorption at -62 km~s$^{-1}$ where there appears prominent emission. 
Fig. 2 here shows an HI emission peak at -65 km~s$^{-1}$ towards both CTB 37A/B but also clearly shows its respective absorption. 
CTB 37A's CO spectrum also reveals a CO cloud at -65 km~s$^{-1}$ which should partly be associated
 with this HI absorption feature. 
The SGPS data has a sensitivity and (spatial and velocity) resolution much better than the data
 Caswell et al. used. 
Generally, the HI distribution and spectrum has an obvious direction dependence, so high resolution HI data is helpful to construct reliable HI absorption spectrum against a strong background. 
For low resolution HI data, one has to average over large source and background regions to construct HI absorption spectra against a background source, which can lead to loss of some real weak absorption. 
We have also seen the weak -65 km~s$^{-1}$ absorption feature towards G348.5-0.0, this further supports that our detection of the absorption feature is reliable. Table 1 shows the difference between our distance measurements and the past for the three SNRs.

Frail et al. (1996) searched for 1720 MHz OH masers towards 66 SNRs in our Galaxy and detected 
OH maser emission towards CTB 37A/B using the Green Bank 43-m and Parkes 64-m telescopes.
Using follow-up VLA and ATCA observations, they confirmed the detections of compact emission 
from CTB 37A, most of which are between -63.5 and -66.3 km~s$^{-1}$, two are at -21.4 and -23.3 km~s$^{-1}$. 
Most 1720 MHz OH masers have been seen as signposts of SNR-molecular cloud interaction (Yusef-Zadeh et al. 2003), 
so the maser velocities have been used to estimate kinematic distances to CO clouds and associated SNRs. 
Using CO observation of CTB 37A, Reynoso \& Mangum (2000) found most of the masers are located 
inside CO structures at velocity $\sim$ -65 km~s$^{-1}$, which is consistent with the 65 
km~s$^{-1}$ masers' originating from shocked CO clouds.
They obtained a distance estimated of $\sim$ 11.3 kpc for CTB 37A (also they assumed that the OH masers at -21.4 and -23.3 km~s$^{-1}$ are associated with G348.5-0.0 which hints a distance of $\sim$ 14 kpc to the remnant). 
However, they also thought these clouds might not be SNR-shocked gas because their densities of 
10$^{-2}\sim$10$^{-3}$ cm$^{-3}$ are 2 orders of magnitude lower than the predicted value 
(10$^{-5}$ cm$^{-3}$) which is required in order to produce 1720 MHz masers in the shocked cloud.
We concluded above that CTB 37A is inside the 3 kpc ring, so these masers can be 
 caused by the interaction between CTB 37A's shock and some molecular clouds inside 3kpc with
 different random velocities ($\sim$ -23 to -88 km~s$^{-1}$). 
We do not exclude the possibility that some masers might instead be related to star-forming
 regions (Caswell 2004) because the IRAS point source 17111-3824 lies close to the 1720 MHz OH masers at -22 km~s$^{-1}$ (see Fig. 5 of Reynoso \& Mangum 2000). 
At this velociy Hewitt et al. (2008) detected wide absorption for the main line OH masers 
(1665 and 1667 GHz) towards CTB 37A/G348.5-0.0 in the GBT OH maser survey. 
Their GBT observations can not clearly separate the two remnants because of low resolution. 
The main line transitions are usually prominent in massive star-forming regions.   

\subsection{Is SN 393 Associated with the SNR Complex?} 

The SNR complex CTB 37 (including 37A/B and G348.5-0.0) was suggested as counterpart of the 
historical supernova of 393 AD (Clark \& Stephenson 1975). 
Sato et al. (2010) obtained a characteristics age of $\sim$ 1000 years for the magnetar CHOU 
J171405.7-381031 in 37B using $XMM-NEWTON$ observations and considered this as supporting 
evidence for above suggestion. 
Aharonian et al. (2008) used $Chandra$ observations of CTB 37B and estimated its kinematic age 
to be $\sim$ 4900 years (assuming d = 7 kpc) using the Sedov-Taylor phase. 
CTB 37A/B are type II SNe because of the magnetar in 37B and a possible pulsar-wind nebula 
CHOU J171419.8-383023 in CTB 37A (Aharonian et al. 2008, Halpern \& Gotthelf 2010, Sato et al. 2010). 
Type II SNe have a typical maximum-light absolute V magnitude of -17 to -18, so the maximum-light 
V magnitude at 6.3 kpc would be $\sim$-4, uncorrected for extinction. 
For CTB 37A, its visual extinction derived from the X-ray column density of 
N$_{H}\sim$3.2$\times10^{22}$cm$^{-2}$ (Aharonian et al. 2008) is $A_{v} \sim 14$ magnitudes 
(taking N$_{H}\sim 2\times$10$^{21}$ A$_{v}$ cm$^{-2}$ mag$^{-1}$, see Predehl \& Schmitt 1995,
 Gu\"ver \& O\"zel 2009).
The extinction results in a maximum light V magnitude of $\sim$ 10, so that CTB 37A at its
peak would be invisible to the naked eye, and thus not associated with SN 393. 
CTB 37B is also not associated with SN 393 because of its visual extinction and distance are
larger than for CTB 37A. 
Kassim et al. (1991) separated the SNR G348.5-0.0 from CTB 37A using multi-frequency VLA observations. 
G348.5-0.0 is the closest one in the CTB37 SNR complex. 
Non-detection of a neutron star inside G348.5+0.0 hints its progenitor might be a type Ia SN 
with absolute magnitude of -19 to -20. Thus, we cannot exclude the possibility that the 
progenitor of G348.5-0.0 was visible and is associated with the historical event 393 AD. More
detailed observations and determination of the age of G348.5+0.0 are needed to determine if the
association is real.

\section{Acknowledgments}
We thank Dr. Reynoso providing the CO data to us. WWT acknowledges supports from the NSFC (011241001) and BeiRen program 
of the CAS(034031001). DAL receives support from the Natural Sciences and Engineering Research Council of Canada.  

\end{document}